%% file: main.tex
\documentclass[10pt,conference]{IEEEtran}

\usepackage{color}
\usepackage{amsmath}
\usepackage{amsfonts}
\usepackage{algorithm}
\usepackage{algorithmicx}
\usepackage{algpseudocode}
\usepackage{graphicx}

\algnewcommand{\Initialize}[1]{%
  \State \textbf{Initialize:}
  \State \hspace*{\algorithmicindent}\parbox[t]{0.8\linewidth}{\raggedright #1}
}

\title{An Iterative Improvement Method for HHL algorithm for Solving Linear System of Equations}
\author{
\IEEEauthorblockN{Yoshiyuki Saito}
\IEEEauthorblockA{
\textit{Graduate School of Computer Science and Engineering} \\
\textit{University of Aizu}\\
Aizu-Wakamatsu, Fukushima, JAPAN \\
m5241141@u-aizu.ac.jp
}
\and
\IEEEauthorblockN{Xinwei Lee}
\IEEEauthorblockA{
\textit{Graduate School of Systems and Information Engineering} \\
\textit{University of Tsukuba}\\
Tsukuba, Ibaraki, JAPAN \\
xwlee@cavelab.cs.tsukuba.ac.jp
}
\and
\IEEEauthorblockN{Dongsheng Cai}
\IEEEauthorblockA{
\textit{Faculty of Engineering, Information and Systems} \\
\textit{University of Tsukuba}\\
Tsukuba, Ibaraki, JAPAN \\
cai@cs.tsukuba.ac.jp
}
\and
\IEEEauthorblockN{Nobuyoshi Asai}
\IEEEauthorblockA{\textit{School of Computer Science and Engineering} \\
\textit{University of Aizu}\\
Aizu-Wakamatsu, Fukushima, JAPAN \\
nasai@u-aizu.ac.jp
}
}

\begin{document}

\maketitle

\begin{abstract}
	\input{abst.tex}	
\end{abstract}

\section{Introduction}
\input{intro.tex}

\section{The HHL algorithm}
\input{hhl.tex}

\section{Iterative improvement method}
\input{iim.tex}

\section{The Iterative Improvement Method for the HHL algorithm}
\input{iimhhl.tex}

\section{Simulations and Results}
\input{simulation_results.tex}

\section{Conclusion}
\input{conclusion.tex}

\bibliographystyle{IEEEtran}
\bibliography{bibfile}

\end{document}

%% file: abst.tex
We propose an iterative improvement method for the Harrow-Hassidim-Lloyd (HHL) algorithm to solve a linear system of equations. This is a quantum-classical hybrid algorithm.
 
The accuracy is essential to solve the linear system of equations.  However, the accuracy of the HHL algorithm is limited by the number of quantum bits used to express the eigenvalues of the matrix. 
Our iterative method improves the accuracy of the HHL solutions, and gives higher accuracy which surpasses the accuracy limited by the number of quantum bits.
 
In practical HHL algorithm, a huge number of measurements is required to obtain good accuracy, even if we provide a sufficient number of quantum bits for the eigenvalue expression, since the solution is statistically processed from the measurements.
Our improved iterative method can reduce the number of measurements.
 
Moreover, the sign information for each eigenstate of the solution is lost once the measurement is made, although the sign is significant.
Therefore, the naïve iterative method of the HHL algorithm may slow down, especially, when the solution includes wrong signs.

In this paper, we propose and evaluate an improved iterative method for the HHL algorithm that is robust against the sign information loss, in terms of the number of iterations and the computational accuracy.

%% file: intro.tex
In this paper, we consider a linear system of equations $Ax = b$ where $A \in \mathbb{R}^{N\times N}$ be Hermitian matrix and $b\in \mathbb{R}^N$.

The linear system of equations is essential in a wide range of fields such as science and engineering, and we have to solve a linear system of equations both at high speed and with high accuracy.
To solve it with high speed and accuracy, we focus on the structure of the matrix $A$ (dense, sparse, symmetric, and etc.).
In general, there are two types of algorithms for solving linear system of equations: iterative methods, represented by the conjugate gradient method \cite{hestenes1952methods}, and direct methods, represented by the Gauss elimination method \cite{golub2013matrix}.
In the iterative method, the solution converges so that the evaluation function such as the residual converge to 0.
In the direct method, the solution can be obtained in the finite number of operations.
The accuracies of both methods are influenced by rounding errors.
However, since, in the iterative method, its calculations repeat until the residual to be zero, it can improve the accuracy even if it is influenced by the rounding-off errors.
On the other hands, since the direct method ends after a finite number of operations, further improvement in accuracy is not possible.
Fortunately, with the iterative improvement method in addition to the direct method allows to improve the accuracy \cite{wilkinson1994rounding}.
The iterative improvement method \cite{wilkinson1994rounding} aims to improve the solution accuracy by iterative calculation in addition to the direct method.
We want to try to improve the accuracy through various methods.

On the other hands, in quantum computings,  
a quantum algorithm for a linear system of equations entitled the HHL algorithm is proposed by Harrow et al. \cite{harrow2009quantum}.
The HHL algorithm is exponentially faster than a classical algorithm to solve a linear system of equations with sparse Hermitian matrices\cite{dervovic2018quantum}.
In addition, the HHL algorithm has been applied to the least squares method \cite{PhysRevLett.109.050505} and supervised machine learning \cite{lloyd2013quantum} due to the potential speed up in quantum computers.
In practical applications, when the HHL algorithm is used as a subroutine, the computational accuracy of the whole system can be limited by the HHL algorithm. Therefore, we should pay attention  not only on the run time but also on the estimated accuracy of the solution.
The estimated accuracy of the HHL solution depends on 1) that of the quantum state constructed by the HHL and on 2) the number of measurements.

1) The accuracy of the constructed quantum state is determined both (a) by the computational accuracy of the matrix $e^{iAt}$ in the Hamiltonian simulation with a Hermitian matrix $A$, and (b) by the number of quantum bits to represent the eigenvalues of the matrix $e^{iAt}$ in the Quantum Phase Estimation (QPE) algorithm \cite{berry2007efficient,nielsen2002quantum}.
Since the accuracy of the Hamiltonian simulation is affected by the discretization error in time $t$, we need to use sufficient number of time slices to obtain an enough accuracy \cite{suzuki1990fractal}.
Therefore, we cannot have a higher accurate quantum state than that determined by the given parameters, such as the time slices and the number of quantum bits, once the parameters are fixed.
If the Hamiltonian simulation is accurate enough, we can improve the accuracy by increasing the number of quantum bits for the QPE \cite{Ubbens2019PracticalIO}.
Moreover, the effect of the matrix condition number on the solution accuracy of the HHL algorithm has also been observed by numerical experiment using some $4 \times 4$ Hermitian matrices \cite{ipsj2020saito}.
Consequently, the matrix condition number determines the minimum number of quantum bits for the QPE to obtain the accuracy by one digit.

2) Since the solution is estimated by measurements from the created quantum state, it is necessary to repeat a very large number of measurements to reduce the estimation error.
In addition, the signs of the solution  are important for the accuracy, but it is lost during the measurements.

In this paper, we propose an Iterative Improvement Method for the HHL (IIMHHL) algorithm.
The IIMHHL algorithm is quantum-classical hybrid algorithm. It, respectively, composed with the HHL algorithm for solving a linear system of equations as quantum part and iterative process as classical part.
We can improve the accuracy repeating the iterative improvement method to the HHL algorithm provided that one iteration can improve the accuracy to some extent.
As a consequence, the improved accuracy can exceed the accuracy limit imposed by the number of quantum bits in the QPE.

In the present paper, first, we introduce the HHL algorithm and iterative improvement method.
Second, we explain our iterative improvement algorithm. Third, we evaluate the accuracy of our proposed algorithm using some $4 \times 4$ real Hermitian matrix with the condition numbers $10$ and $100$. Finally, we concluded our IIMHHL algorithm. 

%% file: hhl.tex
First of all, we explain the HHL algorithm.
Let $A$ be a $N \times N$ Hermitian matrix and $b$ be a $N$ dimensional vector.
Here, we try to find a vector $x$ satisfying $Ax = b$.
The HHL algorithm gives a quantum state $|x\rangle = A^{-1}|b\rangle$ from the input quantum state.
The HHL procedure is summarized in Algorithm \ref{alg:HHL}.
\begin{figure}[!t]
\begin{algorithm}[H]
	\caption{The HHL algorithm for solving $A|x\rangle = |b\rangle$}
    \label{alg:HHL}
    \begin{algorithmic}[1]
    	\Require A Hermitian matrix $A$, a state vector $|b\rangle$ and the number of quantum bits $p$ for the QPE.
		\State Prepare the initial state $|b\rangle |0\rangle_p |0\rangle_{a}$.
		\State Apply the Quantum Phase Estimation with $e^{iAt}$,
			$
			|b\rangle |0\rangle_p |0\rangle_{a} 
			\xrightarrow{\text{QPE}} 
			|b\rangle_I |\tilde{\lambda}_j\rangle_p |0\rangle_{a} 
			= \sum_{j=1}^N \beta_j |u_j\rangle|\tilde{\lambda}_j\rangle_p|0\rangle_a
			$.
		\State Apply a controlled rotation on auxiliary bit $|0\rangle_a$ with normalized constant $C$,
		
			$\rightarrow
			\sum_{j=1}^N \beta_j |u_j\rangle|\tilde{\lambda}_j\rangle_p
			\left(\sqrt{1-\frac{C^2}{\tilde{\lambda}_j^2}}|0\rangle_a +\frac{C}{\tilde{\lambda}_j}|1\rangle_a \right)$.
			
		 \State Uncompute $|\tilde{\lambda}_j\rangle_p$
		
			$\rightarrow
			\sum_{j=1}^N \beta_j |u_j\rangle|0\rangle_p
			\left(\sqrt{1-\frac{C^2}{\tilde{\lambda}_j^2}}|0\rangle_a +\frac{C}{\tilde{\lambda}_j}|1\rangle_a \right)$.
		
		\State Observe auxiliary bit,
			
			If $|0\rangle_a$ is measured, goto step 1.
			
			If $|1\rangle_a$ is measured, observe the state			
			$|x\rangle = \sum_{j=1}^N \frac{C\beta_j}{\tilde{\lambda}_j} |u_j\rangle = A^{-1}|b\rangle$ and record the measured state.
		\State Repeat as many times as needed.
		
		\State The solution from statistical processing $x_{sta}$.
		
		\State \textbf{Post process}
		
			Scale factor $f_1 = \frac{||b||_2}{||Ax_{sta}||_2}$ .
			
			Rotation angle $f_2 = \arg(b\cdot Ax_{sta})$.
			
		\Ensure The approximate solution $x = f_1 e^{if_2} x_{sta}$.
\end{algorithmic}
\end{algorithm}
\end{figure}
Here, $|b\rangle$ is a quantum state corresponding to vector $b$.
Since the eigenvalue of the unitary matrix is $e^{i2\pi\theta},0\leq \theta < 1$, in order to maintain the correspondence between the eigenvalue $\lambda_j$ of the Hermitian matrix $A$ and the eigenvalue $e^{i\lambda_j t}$ of the unitary matrix $e^{iAt}$, we need to use the parameter $t$ that maps all eigenvalues of the matrix $A$ to the interval $[0, 2\pi)$.
If $|u_j\rangle$ is assumed to be the eigenvector corresponding to the eigenvalue $\lambda_j$ of the matrix $A$, then $|b\rangle$ is represented by the linear combination of $|u_j\rangle$ and $\beta_j$, where $\beta_j$ is the coefficient of expanded terms.
Here, $|\tilde{\lambda}_j\rangle_p$ is a quantum state $|l_{j,1}l_{j,2}\cdots l_{j,p}\rangle_p$ where $l_{j,1}l_{j,2}\cdots l_{j,p}$ is the first $p$ bits of the binary representation of $\lambda_j t = 2\pi 0.l_{j,1}l_{j,2}\cdots l_{j,p}\cdots$.

Now, we have to note that the accuracy of $|x\rangle$ depends on the quantum bits $p$ assigned to the QPE to represent eigenvalues.
Since the upper limit of the accuracy of a generated quantum state is determined by the number of quantum bits $p$, we can generate a quantum state with higher accuracy by increasing the number of quantum bits $p$. 
We can estimate the better solution $x_{sta}$ with higher accuracy both by generating a quantum state with higher precision and measuring it many times.
In other words, we can not obtain the solution with higher accuracy without the quantum state with higher accuracy.

Since the solution $x_{sta}$ estimated by the HHL algorithm is normalized, the size of $x_{sta}$ is scaled to that of the original solution by $f_1$.
Although the sign information of each element of the solution $x_{sta}$ is important to estimate its accuracy, it is lost during statistical processing.
Instead of the sign of each element, we use the sign $e^{if_2}$ as the solution vector and evaluate the accuracy using $x = f_1 e^{if_2} x_{sta}$.

For any matrix $A$,
\begin{equation}
	\tilde{A} := 
	\begin{bmatrix}
		0 & A \\
		A^* & 0
	\end{bmatrix}
\end{equation}
is Hermitian matrix and 
\begin{equation}
	\tilde{A}\tilde{x} = \tilde{b}
\end{equation}
would be suitable for the HHL algorithm, where 
$\tilde{x} = \begin{bmatrix}
	0 \\ x
\end{bmatrix}$,
$
\tilde{b} = \begin{bmatrix}
	b \\ 0 
\end{bmatrix}
$.

%% file: iim.tex
The iterative improvement method for a linear system of equations \cite{wilkinson1994rounding} is a practical technique.
This method update to the solution both (1) by calculating the residual from the previous solution, and (2) by solving the new equation applying the residual to the right-hand side.
By updating the solution, we can obtain a solution with higher accuracy than the previous one.
The iterative improvement algorithm is given as Algorithm \ref{alg:IIM}.
\begin{figure}[!t]
\begin{algorithm}[H]
	\caption{Iterative improvement method for solving $Ax = b$}
	\label{alg:IIM}
	\begin{algorithmic}[1]
    	\Require A matrix $A$, a vector $b$.
    	\Initialize{
    		$r_0 = b$,
    		$x = 0$
    	}
     	\For{$m=0, 1, 2,\ldots$}
       		\State Solve $Ay_m = r_m$.
    		\State Update $x \leftarrow x + y_m$.
      		\State Compute residual $r_m = b - Ax$.
    	\EndFor
     	\Ensure $x$
	\end{algorithmic}
\end{algorithm}
\end{figure}

Unfortunately, the residual calculation requires $\mathcal{O}(N^2)$ time complexity or computation.
If a matrix $A$ is sparse, it requires $\mathcal{O}(N)$ to calculate the residual.

%% file: iimhhl.tex
Our proposed Iterative Improvement Method for the HHL (IIMHHL) algorithm is an iterative improvement method (Algorithm \ref{alg:IIM}) using the HHL algorithm (Algorithm \ref{alg:HHL}) as the linear solver engine, for example, Gaussian elimination etc..
We use a quantum state $|v\rangle$ corresponding to a vector $v$. 
Our iterative procedure is described as Algorithm \ref{alg:IIMHHL}.
\begin{figure}[H]
\begin{algorithm}[H]
	\caption{Iterative improvement method for the HHL algorithm}
	\label{alg:IIMHHL}
	\begin{algorithmic}[1]
    	\Require A matrix $A$, a vector $b$ and the number of quantum bits $p$ for the QPE.
		\Initialize{
		$r_0 = b, \tilde{x}_0 = 0, x = 0$.
		}
		
		\For{$m=0, 1, 2, \ldots$}
			\State Solve $A|y_m\rangle = |r_m\rangle$ by the HHL and get the approximate solution $y_{m}$.
			\State $x_m = y_m - \tilde{x}_m$.
			\State Update $x \leftarrow x + x_m$.
			\State Determine a shift vector $\tilde{x}_{m+1}$
			\State Compute residual $r_{m+1} = b - A(x - \tilde{x}_{m+1}) $
		\EndFor
     	\Ensure $x$
	\end{algorithmic}
\end{algorithm}
\end{figure}

The iterative improvement method leads to a more accurate solution based on the first solution.
However, since the signs of each vector components of the obtained solution $y_m$ by the HHL are all the same, the na\"{i}ve iterative improvement method using only $y_m$ may slow down the error convergence if the solution vector components contain different signs.
We can control the convergence speed of the error using the shift vector $\tilde{x}_m$ to make all solution vector components to be positive by shifting operations.
In iteration $m=1,2,\ldots$, the solution vector includes the shift vector $\tilde{x}_m$, and the update vector becomes $y_m - \tilde{x}_m$.

There is a limit to reduce the error in the iterative improvement method.
Since our iterative improvement method using the HHL algorithm is a hybrid method that include a classical iterative process, the maximum accuracy that can be improved is limited by the number of significant digits in a classical computer.
In this paper, because we perform our hybrid algorithm in double precision floating point, the number of significant digits in its precision is about 15 digits in decimal representation \cite{goldberg1991every}.

%% file: simulation_results.tex
In order to evaluate the accuracy of solution by the iterative improvement method for the HHL algorithm, we use $4 \times 4$ matrices to solve the linear system fo equations.
We use the Qiskit\cite[ver 0.22]{Qiskit} to simulate the HHL algorithm that is quantum part of the our proposed method.
Using the Qiskit, we can solve linear systems of equations using state vector and measurements emulating a quantum computer.
We simulate quantum part under ideal conditions without noise using the Qiskit.

\subsection{Problem setting}
In order to consider the influence of the condition number and the signs of the solution vector components,  we set up the matrix $A$ and the vector $b$ as follows.
A Hermitian matrices with condition number $\kappa = 10, 100$ are constructed by similarity transformation of a diagonal matrix with maximum eigenvalue $1$ and minimum eigenvalue $1/\kappa$ by real unitary matrices $U_k,(k=1,2)$,
\begin{equation}
  A_{\kappa,k} = U_k^* \text{diag}[1, 0.5, 0.1, 1/\kappa] U_k.
\end{equation}
Now, we consider the Hermitian matrix so the condition number $\kappa$ can be defined by the ratio of the largest eigenvalue $\lambda_{max}$ to the smallest eigenvalue $\lambda_{min}$, $\kappa = \frac{|\lambda_{max}|}{|\lambda_{min}|}$.
In order to consider the influence of the condition number $\kappa$, we created the matrix $A_{\kappa, k}$ using the same unitary matrix $U_k$.
In other words, $A_{10,k}$ and $A_{100,k}$ are made by the same unitary matrix $U_k$.
In order to consider the influence of the solution vector, we set a solution $x_1$ in which the signs of the elements are all the same and a solution $x_2$ in which the elements contain different signs:
\begin{equation}
  x_1 = [1, 0.1, 0.01, 10]^T,
\end{equation}
\begin{equation}
  x_2 = [-1, 0.1, 0.01, 10]^T.
\end{equation}

We prepare $b_{\kappa, k}$ by calculating $A_{\kappa, k} x_k,(\kappa=10, 100, k=1, 2)$ in advance, and solved $A_{\kappa, k}x = b_{\kappa, k}$.

\subsection{Method of shifting $\tilde{x}_m$ for the IIMHHL algorithm}
In order to compare the influence of shifting, we test the following five types of shifting $\tilde{x}_m$:
\begin{enumerate}
\item $\tilde{x}_m = [0, 0, 0, 0]^T$  namely, not shift,
\item $\tilde{x}_m = \frac{||x_m||_2}{||x_{m-1}||_2}[1, 1, 1, 1]^T$,
\item $\tilde{x}_m = 0.1 x_{m,abs}$,
\item $\tilde{x}_m = \frac{||x_m||_2}{||x_{m-1}||_2} x_{m,abs}$,
\item $\tilde{x}_m = \sqrt{\frac{||x_m||_2}{||x_{m-1}||_2}} x_{m,abs}$,
\end{enumerate}
where $x_{m, abs} = |x_m|$, taking the absolute values of each element.

The method 1) is not shifting.
The method 2) in which all elements are the same, and the methods 3), 4) and 5) are proportional to the elements of the solution vector $x_m$.

\subsection{Results of the HHL algorithm}
\begin{figure}[!t]
  \centering
  \includegraphics[clip, width=3.5in]{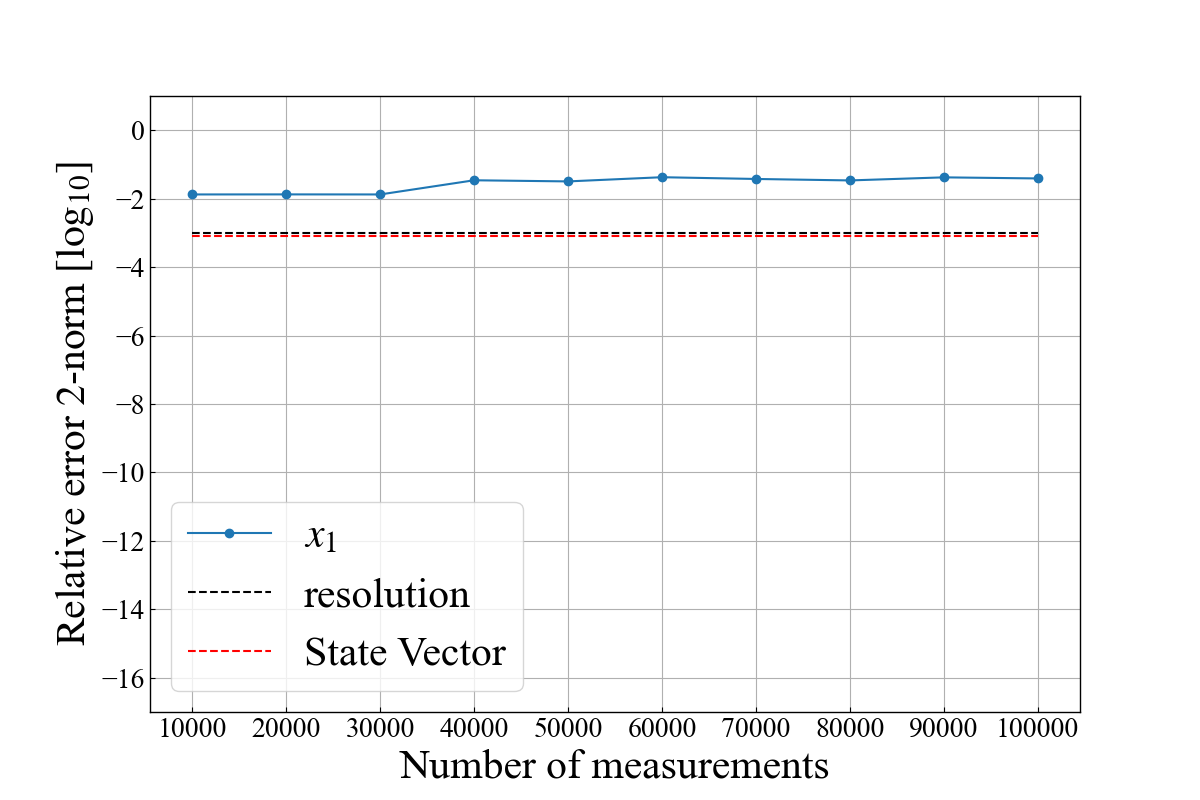}
  \caption{The relative error of the result with $\kappa=10$ and $x_1$ using $p=9$ quantum bits for QPE.
    The black dashed line is the resolution of the $p = 9$ quantum bits.
  }
  \label{fig:many_shots}
\end{figure}

Figure \ref{fig:many_shots} shows the relative error of the solution $x$ of $A_{10,1}x = b_{10,1}$ by the HHL algorithm with $p=9$ quantum bits for the QPE and $6$ time slices for Hamiltonian simulation.
The value obtained by removing the sign from the vertical axis is the number of digits of the precision.
The black dashed line is the resolution of the $p=9$ quantum bits for the QPE.
The red dashed line is the relative error of the solution calculating the state vector.
The calculation by the state vector can be calculated to the accuracy by the quantum bits $p=9$.
The simulated solution accuracy is saturated even if we use many measurements so we can not obtain higher accuracy by the HHL algorithm.

\subsection{Results of the Iterative Improvement Method for the HHL algorithm}
In numerical simulations, we fix the time slices for Hamiltonian simulation to $6$.
When we calculated the matrix of $\kappa = 10$, we used $4$ quantum bits for the QPE and $1,000$ times measurements for each iteration, and,
when we calculated the matrix of $\kappa = 100$, we used $7$ quantum bits for the QPE and $10,000$ times measurements for each iteration.
In addition, we calculated 20 iterations.

\begin{figure}[!t]
  \centering
  \includegraphics[clip, width=3.5in]{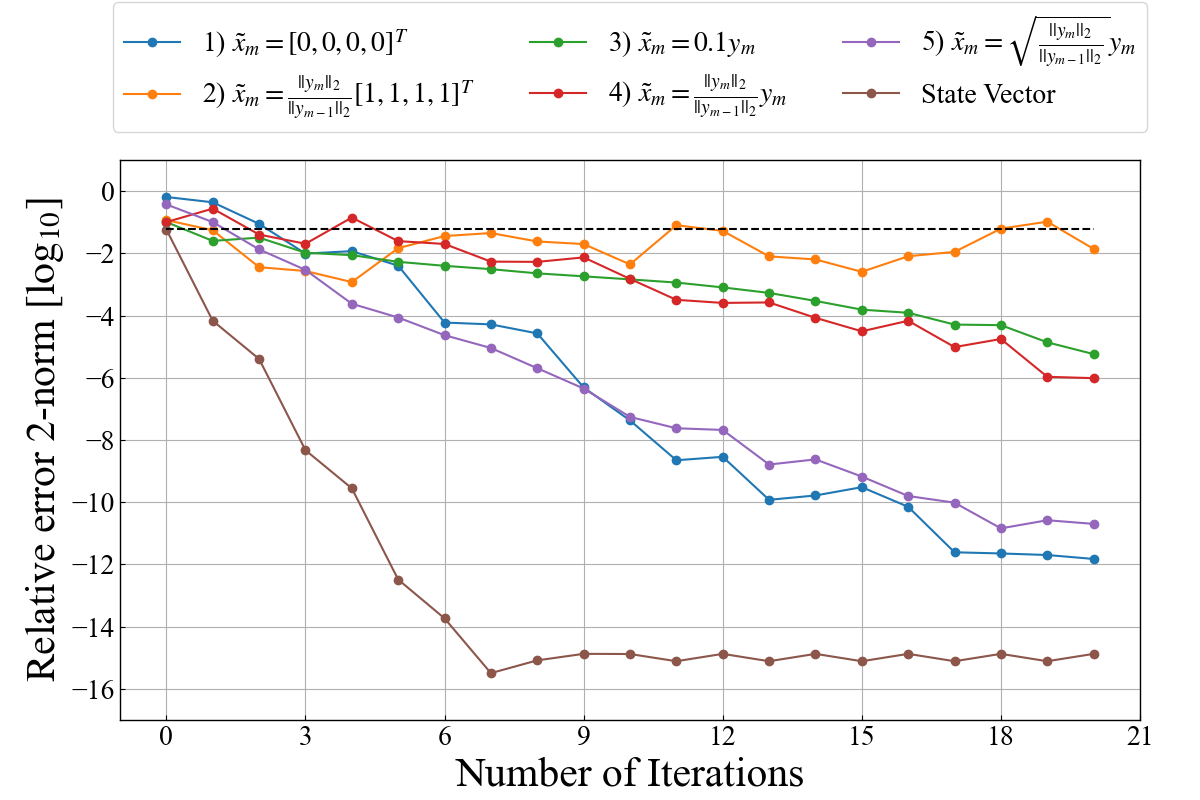}
  \caption{The relative error of the result with $\kappa=10$ and $x_1$ using $p=4$ quantum bits with 1000 measurements for each iteration. The black dashed line is the resolution.}
  \label{fig:x1_10_3_clock4_shots1000_t6}
\end{figure}
Figure \ref{fig:x1_10_3_clock4_shots1000_t6} shows the errors of the solution obtained by the IIMHHL of the linear equation $A_{10,1}x = b_{10,1}$ with condition number $10$ and solution vector $x_1$.
Since the state vector calculation can use the pre-measured quantum state as a solution, we obtain the best result, and the accuracy achieved to the upper limit of the accuracy on the classical computer with $7$ iterations.
In the HHT calculation results by measurement, the method without shifting 1) had the best accuracy or converge faster than other shifting method, obtaining an accuracy of 12 digits.
Method 5) was the second most accurate with a 10 digits precision.
Methods 3) and 4) decrease the error at the same speed, and method 2) did not improve the accuracy due to the oscillation of the error after repeated iterations.

\begin{figure}[!t]
  \centering
  \includegraphics[clip, width=3.5in]{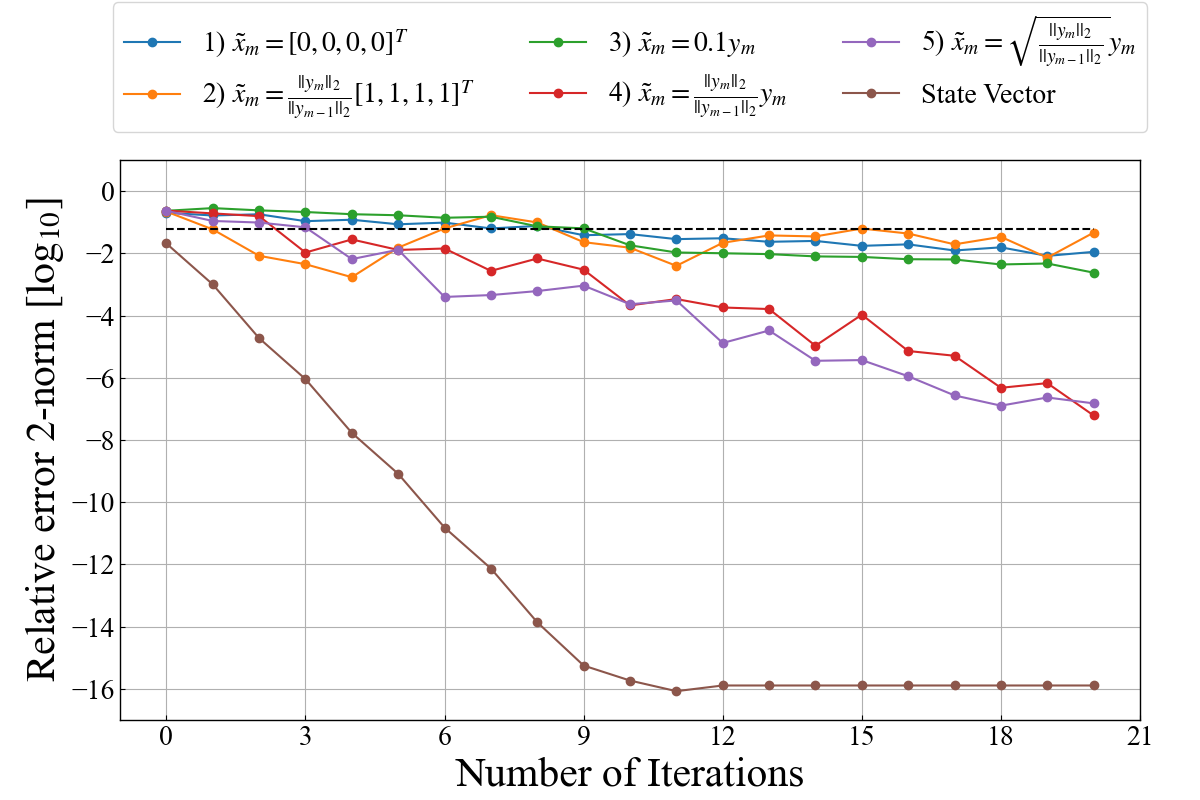}
  \caption{The relative error of the result with $\kappa=10$ and $x_2$ using $p=4$ quantum bits with 1000 measurements for each iteration. The black dashed line is the resolution.}
  \label{fig:x2_10_3_clock4_shots1000_t6}
\end{figure}
Figure \ref{fig:x2_10_3_clock4_shots1000_t6} shows the solution errors  obtained by the IIMHHL for the linear system of equation $A_{10,2}x = b_{10,2}$ with condition number $10$ and solution vector $x_2$.
Since the calculation by the state vector can maintain the sign information, the error falls monotonically by iteration even if different signs were mixed in the solution, and we obtain the 16 digits accuracy.
Methods 4) and 5) had a 7 digits accuracy and the error reduction speeds is the same.
Method 1) had 1 digit accuracy with 20 iterations and the improvement speed is very slow.
Method 3) slowly improve the accuracy.
Method 2) do not improve the accuracy even if 20 iterations are performed.
Methods 4) and 5) can improve the accuracy even if the solution vector contains components with different signs.
However, there are some methods such as methods 1) and 3), where the improved accuracy speed is extremely slow.

\begin{figure}[!t]
  \centering
  \includegraphics[clip, width=3.5in]{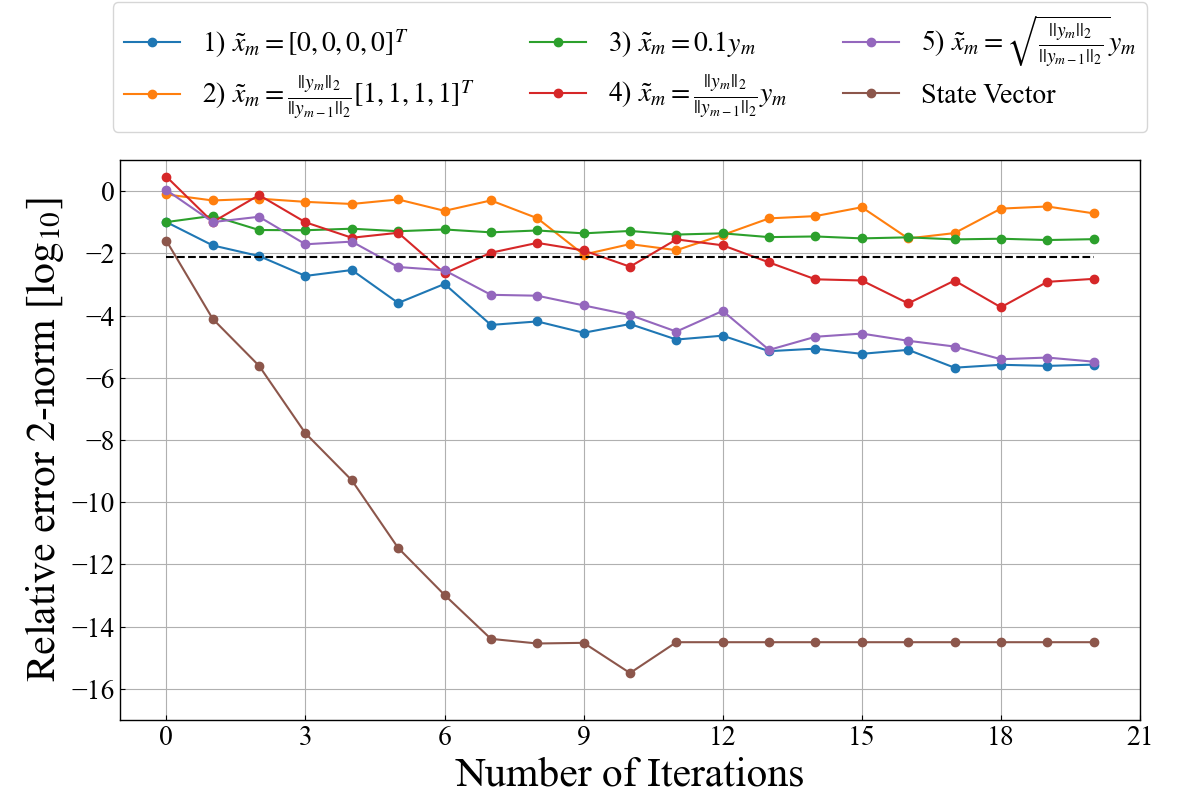}
  \caption{The relative error of the result with $\kappa=100$ and $x_1$ using $p=7$ quantum bits with 10000 measurements for each iteration. The black dashed line is the resolution.}
  \label{fig:x1_100_3_clock7_shots10000_t6}
\end{figure}

Figure \ref{fig:x1_100_3_clock7_shots10000_t6} shows the solution errors obtained by the IIMHHL for the linear system of equation $A_{100,1}x = b_{100,1}$ with condition number $100$ and solution vector $x_1$.
The state vector calculation was able to improve the accuracy up to the precision limit even if the condition number is increased.
Methods 1) and 5) improve the accuracy up to about $4$ digits.
Method 4) can improve the accuracy $2$ digits while the error oscillated.
Methods 2) and 3) cannot improve the accuracy.
Comparing with fig. \ref{fig:x1_10_3_clock4_shots1000_t6}, we observe that the speed of error convergence becomes slower as the condition number increases.

\begin{figure}[!t]
  \centering
  \includegraphics[clip, width=3.5in]{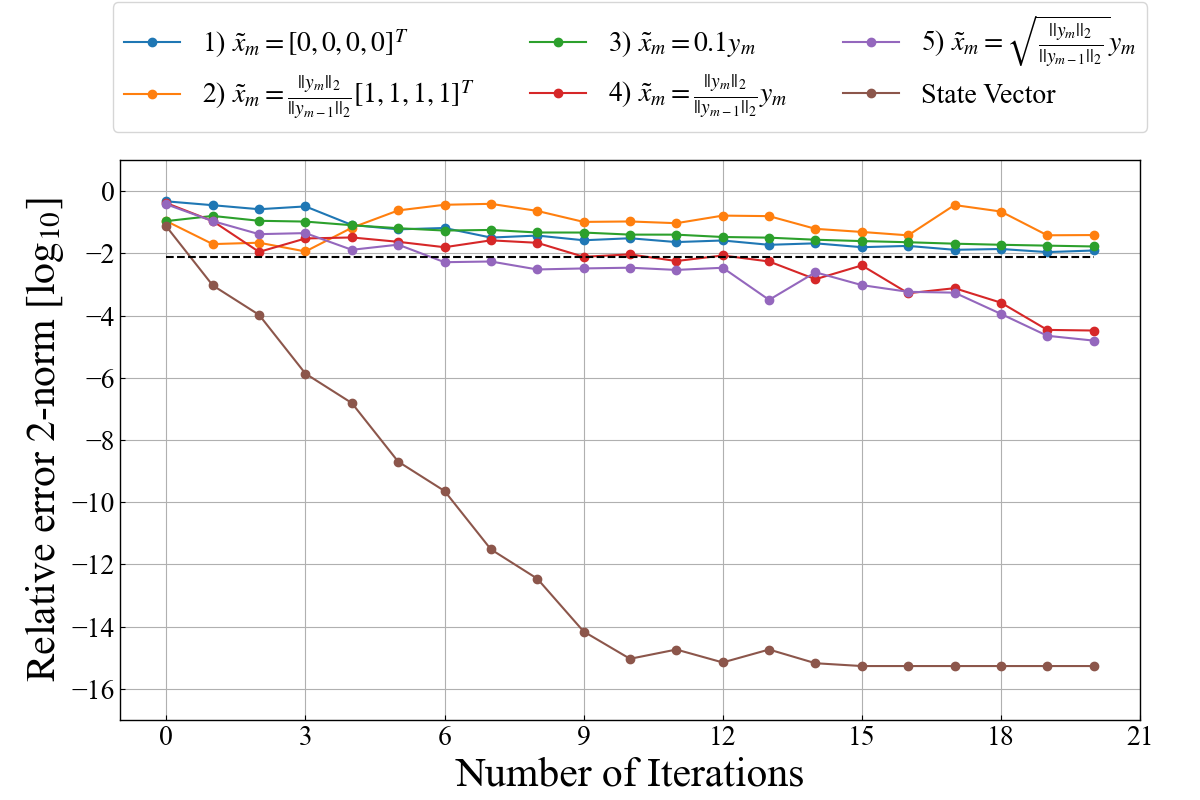}
  \caption{The relative error of the result with $\kappa=100$ and $x_2$ using $p=7$ quantum with 10000 measurements for each iteration. The black dashed line is the resolution.}
  \label{fig:x2_100_3_clock7_shots10000_t6}
\end{figure}
Figure \ref{fig:x2_100_3_clock7_shots10000_t6} shows the errors of the solution obtained by the IIMHHL for the linear system of equation $A_{100,2}x = b_{100,2}$ with condition number $100$ and solution vector $x_2$.
The state vector calculation can improve the accuracy up to the limit even if the condition number increases and the solution contains the mixed signs.
Methods 4) and 5) improve the accuracy $4$ digits, albeit slowly.
Methods 1) and 3) were able to improve the accuracy $1$ digit, albeit very slowly.
Method 2) was cannot improve the accuracy.
Comparing with fig. \ref{fig:x2_10_3_clock4_shots1000_t6}, we can see that the error convergence speed becomes also slower as the condition number increases.

\section{Discussion and Further experiments}
We proposed an iterative improvement method for the HHL algorithm.
This is one of numerical methods to calculate real numerical numbers using a quantum computer.

Since the results of the HHL algorithm are obtained by statistical processing from observation, we can not obtain the sign information accurately if vector components in the solution vector contain different signs, but if the solution elements are all positive or negative, we can adjust the signs by $e^{if_2}$.
If the signs of the solution elements are the same, we can improve the accuracy by the iterative improvement method without shifting.
In fact, method 1) of figs. \ref{fig:x1_10_3_clock4_shots1000_t6} and \ref{fig:x1_100_3_clock7_shots10000_t6} show that we can improve the accuracy  without shifting if the components are the same.
In contrast, figs. \ref{fig:x2_10_3_clock4_shots1000_t6} and \ref{fig:x2_100_3_clock7_shots10000_t6} show that the accuracy improvement speed slows down if the solution vector components have mixed signs and that we can improve the accuracy by shifting properly.
In general, the solution vector components of the linear system of equations contain different signs.
Therefore, na\"{i}ve iterations without shifting may not yield the expected solution even after repeated iterations.
The fact that the accuracy can be improved by iterations of shifting means that our proposed IIMHHL recovers the sign information of each element of the solution vector and is robust against missing sign information.
Furthermore, we performed the shift operation expecting that all the solution vector components are positive, but in fact, not all of them have positive signs, so the error increases when we update the solution.
Nevertheless, the fact that the accuracy can be further improved by iteration confirms that the IIMHHL is robust against missing sign information.
Figures \ref{fig:x2_10_3_clock4_shots1000_t6} and \ref{fig:x2_100_3_clock7_shots10000_t6} show that no shifting method 1) can not improve the accuracy if the solution includes different signs in the vector components.

We performed an additional experiment to confirm whether the shift preprocessing, which makes all solution components positive, can improve the accuracy by iterations as shown in figs. \ref{fig:x1_10_3_clock4_shots1000_t6} and \ref{fig:x1_100_3_clock7_shots10000_t6} even with the no shifting method 1).

\begin{figure}[!t]
  \centering
  \includegraphics[clip, width=3.5in]{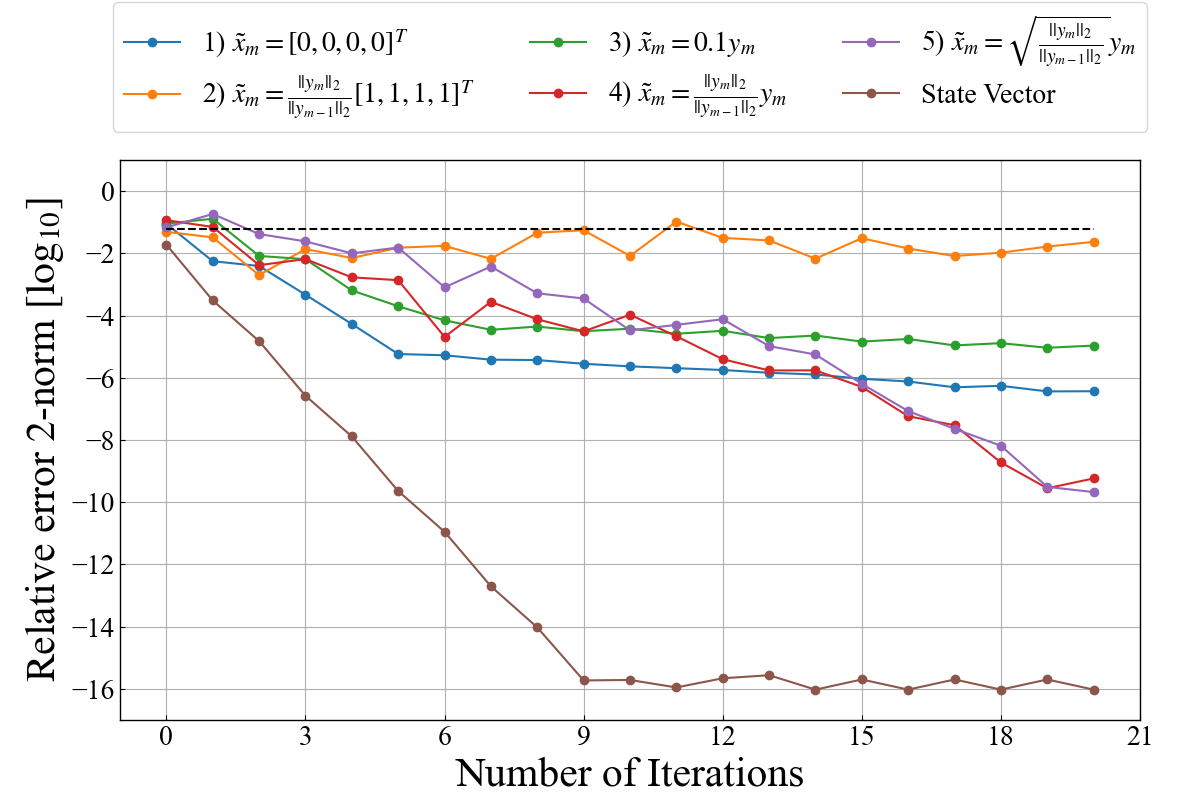}
  \caption{The relation between the relative error, the number of measurements and iterations with $\kappa = 10$ and $x_2 + [2,2,2,2]^T$ using $p=4$ quantum bits for QPE.
    The number of dots is the number of iterations.
  }
  \label{fig:pre_shift}
\end{figure}
Figure \ref{fig:pre_shift} shows the result of the calculation of the linear equation $A_{10,2}x = b_{10,2} + A[2,2,2,2]^T$ after preprocessing to shift by $[2,2,2,2]^T$ so that all components of the solution vector $x_2$ are positive with the condition number $10$.
As the preprocess, shifting all solution components to positive, methods 4) and 5) can obtain better accuracy than the results of fig. \ref{fig:x2_10_3_clock4_shots1000_t6}.
In addition, methods 1) and 3) can also obtain better accuracy than fig. \ref{fig:x2_10_3_clock4_shots1000_t6}, but the accuracy improvement speed slows down in the middle of the process.
Method 2) can not improve the accuracy as good as fig. \ref{fig:x2_10_3_clock4_shots1000_t6}.
Therefor, even if the solution vector components have different signs, the error can be reduced without shifting during the iterative calculation by preprocessing shifting.
However, since the error may decrease slowly during the iteration, it is necessary to find appropriate shifts in each iteration to obtain a faster convergence.

Considering from others aspect of the accuracy, our method can reduce the number of measurements.
In order to obtain each element of the $N$-dimensional solution vector from the HHL algorithm with $d$ digit accuracy, we consider that a sufficient number of quantum bits and $\mathcal{O}(N10^d)$ number of measurements are required to make the quantum state with $d$ digit accuracy.
Therefore, we need $\mathcal{O}(N10)$ measurements to gain $1$ digit accuracy and the number of quantum bits to make a quantum state with $1$ digit accuracy.
We can obtain $d$ digit accuracy by improving $1$ digit accuracy $d$ times by iterative improvement method and using only $d \times \mathcal{O}(N10) = \mathcal{O}(Nd10)$ measurements.
In other words, the number of measurements can be reduced from the order of exponential power of the required accuracy to the order of polynomials.
This might be a solution to reduce the huge number of measurements that are expected in getting real number solution using a quantum computer.
Furthermore the number of quantum bits for the Quantum Phase Estimation algorithm can also be reduced since it only needs to be accurate enough to drop the error by the iterative method, thus reducing the depth of the quantum circuit of the HHL algorithm.

\begin{figure}[!t]
  \centering
  \includegraphics[clip, width=3.5in]{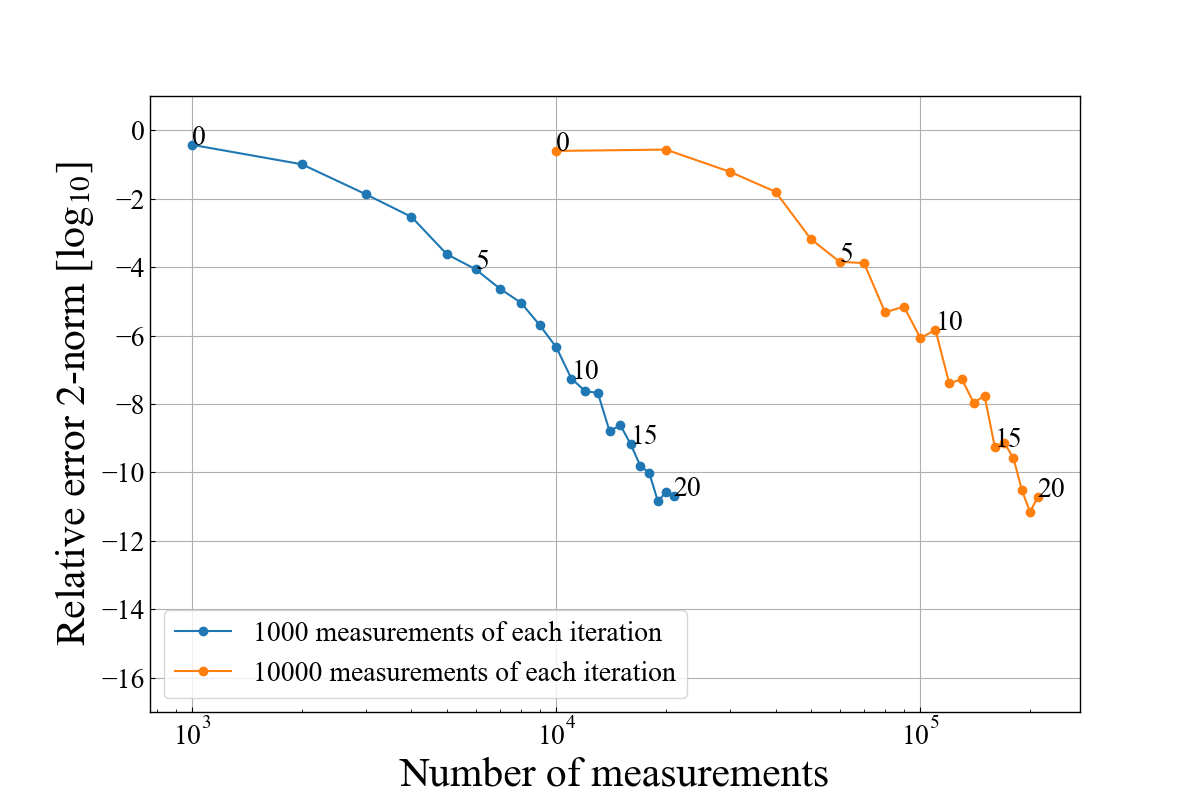}
  \caption{The relation between the relative error, the number of measurements and iterations with $\kappa = 10$ and $x_1$ using $p=4$ quantum bits for QPE and the shift method 5).
    The number of dots is the number of iterations.
  }
  \label{fig:total_shots}
\end{figure}
Figure \ref{fig:total_shots} is an excerpt of the result of shift method 5) from fig.\ref{fig:x1_10_3_clock4_shots1000_t6}, and the result of calculation with the same conditions except that only the number of measurements in each iteration is set to $10,000$.
Here the horizontal axis is the total number of measurements.
This figure show the relationship between the error, the number of measurements, and the number of iterations.
Looking at the number of measurements of $10^4$, we can see that the result of $10$ iteration of $1,000$ measurements HHL achieve the 6 digits accuracy.
A solution with higher accuracy can be obtained with $1,000\times 10 = 10^4$ measurements.
However, if we look at the results of the iterations with $10^4$ measurements, we have not been able to obtain even one digit of accuracy when $10^4$ measurements are performed.
Consequently, simple HHL algorithm with $10,000$ measurements get less  1 digit.
In other words, by increasing the number of iterations instead of increasing the number of measurements, we can obtain a higher-accurate solution with the same number of measurements as a whole.

%% file: conclusion.tex
We have proposed and tested an iterative improvement method for the HHL algorithm to solve some small linear system of equations.
Our proposed method can solve the linear equation with higher accuracy even if the solution vector components contain different signs.
In addition, we can speed up the error convergence by taking  a shift during the iteration to make the solution element positive.
However, the error convergence speed still slows down compared to the state vector calculations, which means that the convergence speed needs to be improved.

We have also demonstrated that our IIMHHL method has a possibility to decrease the number of measurements for expecting continuous value of solution with higher accuracy.
Similar technique may be applicable to other application if an arbitrary function for iteration can be defined.